\documentclass{article}

\def \be {\begin{equation}}
\def \ee {\end{equation}}
\def \bea {\begin{eqnarray}}
\def \eea {\end{eqnarray}}
\def \nn {\nonumber}

\def \a {\alpha}
\def \b {\beta}

\def \G {\Gamma}
\def \d {\delta}

\def \m {\mu}
\def \n {\nu}
\def \k {\kappa}

\def \s {\sigma}
\def \r {\rho}
\def \o {\omega}

\def \th {\theta}
\def \Th {\Theta}

\def \t {\tau}
\def \dag {\dagger}
\def \p {\partial}

\def\bd{\begin{document}}
\def\ed{\end{document}}
\def\nn{\nonumber}
\def\bea{\begin{eqnarray}}
\def\eea{\end{eqnarray}}
\let\bm=\bibitem
\let\la=\label

\def\N{{\cal N}}
\def\sst{\scriptscriptstyle}
\def\thetabar{\bar\theta}
\def\Tr{{\rm Tr}}
\def\one{\mbox{1 \kern-.59em {\rm l}}}

\def\a{\alpha}      \def\da{{\dot\alpha}}
\def\b{\beta}       \def\db{{\dot\beta}}
\def\c{\gamma}  \def\C{\Gamma}  \def\cdt{\dot\gamma}
\def\d{\delta}  \def\D{\Delta}  \def\ddt{\dot\delta}
\def\e{\epsilon}        \def\vare{\varepsilon}
\def\f{\phi}    \def\F{\Phi}    \def\vvf{\f}
\def\h{\eta}
\def\k{\kappa}
\def\l{\lambda} \def\L{\Lambda}
\def\m{\mu} \def\n{\nu}
\def\o{\omega}
\def\P{\Pi}
\def\r{\rho}
\def\s{\sigma}  \def\S{\Sigma}
\def\t{\tau}
\def\th{\theta} \def\Th{\Theta} \def\vth{\vartheta}
\def\X{\Xeta}
\def\z{\zeta}
\def\w{\wedge}
\def\u{\underline}
\def\hs{\hspace}

\def\cA{{\cal A}} \def\cB{{\cal B}} \def\cC{{\cal C}}
\def\cD{{\cal D}} \def\cE{{\cal E}} \def\cF{{\cal F}}
\def\cG{{\cal G}} \def\cH{{\cal H}} \def\cI{{\cal I}}
\def\cJ{{\cal J}} \def\cK{{\cal K}} \def\cL{{\cal L}}
\def\cM{{\cal M}} \def\cN{{\cal N}} \def\cO{{\cal O}}
\def\cP{{\cal P}} \def\cQ{{\cal Q}} \def\cR{{\cal R}}
\def\cS{{\cal S}} \def\cT{{\cal T}} \def\cU{{\cal U}}
\def\cV{{\cal V}} \def\cW{{\cal W}} \def\cX{{\cal X}}
\def\cY{{\cal Y}} \def\cZ{{\cal Z}}

\def\ua{\underline{\alpha}} \def\ubb{\underline{\beta}}
\def\ug{\underline{\gamma}}
\def\ub{\underline{\phantom{\alpha}}\!\!\!\beta}
\def\uc{\underline{\phantom{\alpha}}\!\!\!\gamma}
\def\um{\underline{\mu}} \def\un{\underline{\nu}}
\def\ud{\underline\delta}
\def\ue{\underline\epsilon}
\def\una{\underline a}\def\unA{\underline A}
\def\unb{\underline b}\def\unB{\underline B}
\def\unc{\underline c}\def\unC{\underline C}
\def\und{\underline d}\def\unD{\underline D}
\def\une{\underline e}\def\unE{\underline E}
\def\unf{\underline{\phantom{e}}\!\!\!\! f}\def\unF{\underline F}
\def\unm{\underline m}\def\unM{\underline M}
\def\unn{\underline n}\def\unN{\underline N}
\def\unp{\underline{\phantom{a}}\!\!\! p}\def\unP{\underline P}
\def\unq{\underline{\phantom{a}}\!\!\! q}
\def\unQ{\underline{\phantom{A}}\!\!\!\! Q}
\def\unH{\underline{H}}
\def\ul{\underline}

\def\As {{A \hspace{-6.4pt} \slash}\;}
\def\bs {{b \hspace{-6.4pt} \slash}\;}
\def\Ds {{D \hspace{-6.4pt} \slash}\;}
\def\ds {{\del \hspace{-6.4pt} \slash}\;}
\def\ss {{\s \hspace{-6.4pt} \slash}\;}
\def\ks {{ k \hspace{-6.4pt} \slash}\;}
\def\ps {{p \hspace{-6.4pt} \slash}\;}
\def\pas {{{p_1} \hspace{-6.4pt} \slash}\;}
\def\pbs {{{p_2} \hspace{-6.4pt} \slash}\;}

\def\Fh{\hat{F}}
\def\Vh{\hat{V}}
\def\Xh{\hat{X}}
\def\ah{\hat{a}}
\def\xh{\hat{x}}
\def\yh{\hat{y}}
\def\ph{\hat{p}}
\def\xih{\hat{\xi}}

\def\psit{\tilde{\psi}}
\def\Psit{\tilde{\Psi}}
\def\tht{\tilde{\th}}

\def\At{\tilde{A}}
\def\Qt{\tilde{Q}}
\def\Rt{\tilde{R}}
\def\Nt{\tilde{N}}

\def\at{\tilde{a}}
\def\st{\tilde{s}}
\def\ft{\tilde{f}}
\def\pt{\tilde{p}}
\def\qt{\tilde{q}}
\def\vt{\tilde{v}}
\def\nt{\tilde{n}}

\def\delb{\bar{\partial}}
\def\bz{\bar{z}}
\def\bD{\bar{D}}
\def\bB{\bar{B}}

\def\bk{{\bf k}}
\def\bl{{\bf l}}
\def\bp{{\bf p}}
\def\bq{{\bf q}}
\def\br{{\bf r}}
\def\bx{{\bf x}}
\def\by{{\bf y}}
\def\bR{{\bf R}}
\def\bV{{\bf V}}

\def\d{\delta}\def\D{\Delta}\def\ddt{\dot\delta}

\def\p{\partial} \def\del{\partial}
\def\xx{\times}
\def\uno{\mbox{1 \kern-.59em {\rm l}}}

\def\trp{^{\top}}
\def\inv{^{-1}}
\def\dag{{^{\dagger}}}

\def\pr{\prime}
\def\rar{\rightarrow}
\def\lar{\leftarrow}
\def\lrar{\leftrightarrow}

\def\q111{$Q^{1, 1, 1}$}

\begin{document}

\title{BPS $M2$-branes in $AdS_4\times Q^{1, 1, 1}$ Dual to Loop Operators}
\author{Jun-Bao Wu$^{\clubsuit}$\footnote{E-mail: wujb@ihep.ac.cn}\hs{2ex} Meng-Qi Zhu$^{\spadesuit}$\footnote{E-mail: mqzfly@pku.edu.cn} \\
\small{$^\clubsuit$Institute of High Energy Physics,
and Theoretical Physics Center for Science Facilities,}\\
\small{ Chinese Academy of Sciences, 19B Yuquan Road,
Beijing 100049, P.R. China}\\
\small {$^\spadesuit$Department of Physics,
and State Key Laboratory of
Nuclear Physics and Technology,}\\
\small{Peking University, 5 Yiheyuan Road, Beijing 100871, P.R. China}\\ }
\date{December 2013}
\maketitle

\abstract{In this paper, we first compute the Killing spinors of $AdS_4\times Q^{1, 1, 1}$ and its certain orbifolds. Based on this, two classes of $M2$-brane solutions
are found. The first class of solutions includes $M2$-branes dual to Wilson loops in the fundamental representation as a special case. The second class includes the candidates
of the holographic description of vortex loops in the dual field theories. }

\section{Introduction}

Many examples about the $AdS_4/CFT_3$ correspondence were established since the seminal paper \cite{ABJM} which itself was inspired by \cite{BL1, BL2, BL3, G1, G2}.
In this correspondence, certain three-dimensional superconformal Chern-Simons-matter theories are proposed to be dual to M-theory on $AdS_4\times X_7$.
The three-dimensional theory has ${\cal N}=1$ ($2, 3$) supersymmetry when $X_7$ is a weak $G_2$ (Sasaki-Einstein, $3$-Sasaki) manifold.
Loop operators play an important role in the studies of this $AdS_4/CFT_3$ duality, as they do in the case of the $AdS_5/CFT_4$ correspondence.
$1/6$-BPS Wilson loops in ABJM theory were first studied in detail in \cite{Drukker, Chen, Rey}. Later a highly non-trivial $1/2$-BPS Wilson loop was constructed in \cite{drukker2}. An interesting explanation on the origin of these Wilson loops  was given in \cite{LeeLee} based on \cite{Berenstein:2008dc}.
Some exact results for Wilson loops were obtained based on powerful tools of supersymmetric localization \cite{Kapustin}. The Wilson loops with quite less supersymmtries
were studied in \cite{Cardinali:2012ru, Kim:2013oza, Marmiroli:2013nza}.

It is certainly interesting to generalize these studies on loop operators to $AdS_4/CFT_3$ correspondences with less supersymmetries since now the dynamics is less
constrained by supersymmetries. In Chern-Simon-matter theories with ${\cal N}=2$ supersymmetries, BPS Wilson loops can be constructed when the loop is a straight line or a circle \cite{GaiottoYin}. This point
is different from the four-dimesional ${\cal N}=1$ gauge theories, although they have the same amount of supercharges.
Half-BPS Wilson loops in generic three-dimensional ${\cal N}=2$ Chern-Simons-matter theories were studied in detail in \cite{Sparks}.  The geometry of the matrix models obtained from localization was connected to the geometry of $M2$-brane slolutions in the holographic description based on results from differential geometry.
There also exists vortex loop, a kind of disordered operator, in these theories. The holographic dual of the vortex loop in ABJM theory was studied in \cite{drukker4}.
The vortex loops in generic ${\cal N}=2$ Chern-Simons-matter theories were studied using localization in \cite{Kapustin2, drukker3} based on \cite{Jafferis:2010un, Hama:2010av}.

The aim of the current paper is to study BPS M2-branes in a concrete example, with duality to loop operators in mind.
The first reason why we picked up the Sasaki-Einstein manifold $Q^{1, 1, 1}$ is that the metric of this manifold is very simple, though its isometry group is small.
 The second, less obvious reason is that the Killing spinor equation is easy to solve on this manifold \footnote{Similar thing was noticed for the five-dimensional Sasaki-Einstein manifold $T^{1, 1}$ \cite{Arean}.}. We further discussed the Killing spinors of certain orbifolds of $AdS_4\times Q^{1, 1, 1}$
by using Lie-Lorentz derivation of spinors with respect to Killing vectors \cite{Kosmann, FigueroaO'Farrill:1999va, Ortin:2002qb}. Based on these results, we found two classes of $M2$-branes. The worldvolumes of these $M2$-branes all have the topology $AdS_2\times S^1$. The $AdS_2$ factor is embedded to the $AdS_4$ part of the background geometry, so these $M2$-branes are candidates for the holographic duals of loop operators. In the first class, the $S^1$ is embedded in $Q^{1, 1, 1}$. This class includes the $M2$-branes dual to Wilson loops in the fundamental representation. We think that our study here is complementary to the results in \cite{Sparks}  based on more abstract mathematical tools. In the second
class of $M2$-branes, this $S^1$ has non-trivial profile in both $AdS_4$ and $Q^{1, 1, 1}$. These $M2$-branes are similar to the $M2$-branes in $AdS_4\times S^7/Z_k$ dual to vortex loops in ABJM theory \cite{drukker4}.

We also noticed that there had been many researches about M-theory on $AdS_4\times Q^{1, 1, 1}$ and its various orbifolds. This is another reason why we choose to study $M2$-branes in this background. Various field theory duals were proposed and checked in \cite{Franco:2008um, Franco:2009sp, Aganagic:2009zk, Benini, Jafferis} \footnote{An old proposal can be found in \cite{Fabbri:1999hw}.}. Localization was performed to obtained a matrix model \cite{Nakwoo1} for the field theory proposed in \cite{Benini, Jafferis}. Superconformal indices were computed in \cite{Cheon:2011th, Eager:2013mua}. Some other membranes and five-branes in this background were studied in \cite{Klebanov:2010tj, Ahn:1999ec, Benishti:2010jn}. Some spinning membranes dual to local operators were found in \cite{Kim}. The Penrose limit of $AdS_4\times Q^{1, 1, 1}$ was studied in \cite{Gursoy:2002tx, Ahn:2002qj}. Some supergravity solutions related to $AdS_4\times Q^{1, 1, 1}$ were discussed in \cite{D6}.

In the next section, we will solve the Killing spinor equations on $AdS_4\times Q^{1, 1, 1}$. Two classes of BPS $M2$-brane solutions will be discussed in section 3.

\section{Killing spinors of $AdS_4\times Q^{1, 1, 1}$}
The metric on $AdS_4\times Q^{1, 1, 1}$ is
\bea ds^2&=&R^2(ds^2_4+ds^2_7),\\
ds^2_4&=&\frac14(\cosh^2u(-\cosh^2\rho dt^2+d\rho^2)+du^2+\sinh^2u d\phi^2),\\
ds^2_7&=&\sum_{i=1}^3\frac18(d\theta^2_i+\sin^2\theta_id\phi_i^2)+\frac1{16}(d\psi+\sum_{i=1}^3\cos\theta_i d\phi_i)^2,\eea
with $\theta_i\in[0, \pi], \phi_i\in [0, 2\pi] \, (i=1, 2, 3), \psi\in [0, 4\pi]$.
The four-form field strength on this background is
\be H_4=\frac{3R^3}8 \cosh^2u\sinh u\cosh\rho dt\w d\rho \w du \w d\phi . \ee
Two kinds of $Z_k$ orbifolds of \q111 were considered in the literature. In the first case \cite{Franco:2008um, Franco:2009sp}, the orbifold is obtained via the identification
$(\phi_1, \phi_2)\sim (\phi_1+\frac{2\pi}{k}, \phi_2+\frac{2\pi}{k})$. In the second case \cite{Aganagic:2009zk}, the identification is $\phi_1\sim \phi_1+\frac{2\pi}{k}$.
We will denote the first orbifold as $Q^{1, 1, 1}/Z_k$ and the second orbifold as $Q^{1, 1, 1}/Z_k^\prime$ from now on.
Flux quantization gives
\bea R&=&2\pi l_p\left(\frac{N}{6vol(Q^{1, 1, 1}/Z_k)}\right)^{1/6}\\
&=&l_p\left(\frac{2^8\pi^2kN}{3}\right)^{1/6},\label{rlp}\eea
where we have used
\be vol(Q^{1, 1, 1}/Z_k)=\frac{\pi^4}{8k}. \ee

In order to find the Killing spinors, we find it very useful to introduce the following one-forms \footnote{Such trick was used for
$T^{1, 1}$ in \cite{Arean}.}
\bea \sigma^1_I&=&d\theta_I, \\
     \sigma^2_I&=&\sin\theta_Id\phi_I,\\
     \sigma^3_I&=&\cos\theta_Id\phi_I,   \eea
with $I=1, 2$ and
\bea w^1&=&-\cos\psi\sin\theta_3d\phi_3+\sin\psi d\theta_3,\\
     w^2&=&\sin\psi\sin\theta_3d\phi_3+\cos\psi d\theta_3, \\
     w^3&=&d\psi+\cos\theta_3d\phi_3,
     \eea
     which satisfy
     \bea d\sigma^i_I+\frac12\epsilon^{ijk}\sigma^j_I\w \sigma^k_I&=&0,\\
         dw^i+\frac12\epsilon^{ijk}w^j\w w^k&=&0. \eea
 Using these one-forms, we can re-express the above metric on \q111 as
 \be  ds^2_7=\sum_{I=1}^2 \frac18[(\sigma^1_I)^2+(\sigma^2_I)^2]+\frac18(w^1)^2+\frac18(w^2)^2+\frac1{16}(\sigma^3_1+\sigma^3_2+w^3)^2. \ee
 Now the vielbeins of the eleven-dimensional metric are
 \bea e^{\ul{0}}&=&\frac{R}2\cosh u\cosh\rho dt, \,
      e^{\ul{1}}=\frac{R}2\cosh u d\rho,\\
      e^{\ul{2}}&=&\frac{R}2 du,\,
      e^{\ul{3}}=\frac{R}2 \sinh u d\phi,\\
      e^{\ul{4}}&=&\frac{R}{2\sqrt{2}} \sigma^1_1,\,
      e^{\ul{5}}=\frac{R}{2\sqrt{2}} \sigma^2_1,\\
      e^{\ul{6}}&=&\frac{R}{2\sqrt{2}} \sigma^1_2,\,
      e^{\ul{7}}=\frac{R}{2\sqrt{2}} \sigma^2_2,\\
      e^{\ul{8}}&=&\frac{R}{2\sqrt{2}} w^1,\,
      e^{\ul{9}}=\frac{R}{2\sqrt{2}} w^2,\\
      e^{\ul{\sharp}}&=&\frac{R}4 (\sigma^3_1+\sigma^3_2+w^3).
 \eea
  The spin connections with respect to these vielbeins are
  \bea \omega^{\ul{0}\ul{1}}&=&\frac{2}{R}\frac{\tanh\rho}{\cosh u}e^{\ul{0}},\,
  \omega^{\ul{0}\ul{2}}=\frac{2}{R}\tanh u e^{\ul{0}},\\
  \omega^{\ul{1}\ul{2}}&=&\frac{2}{R}\tanh u e^{\ul{1}},\,
  \omega^{\ul{2}\ul{3}}=-\frac{2}{R} \coth u e^{\ul{3}},\\
  \omega^{\ul{4}\ul{5}}&=&\frac{1}{R}(-2\sqrt{2}\cot\theta_1 e^{\ul{5}}+e^{\ul{\sharp}}),\,
  \omega^{\ul{6}\ul{7}}=\frac{1}{R}(-2\sqrt{2}\cot\theta_2 e^{\ul{7}}+e^{\ul{\sharp}}),\\
  \omega^{\ul{8}\ul{9}}&=&\frac{1}{R}(2\sqrt{2}\cot\theta_1 e^{\ul{5}}+2\sqrt{2}\cot\theta_2 e^{\ul{7}}-3e^{\ul{\sharp}}),\\
 \omega^{\ul{4}\ul{\sharp}}&=&\frac{1}{R}e^{\ul{5}}, \omega^{\ul{5}\ul{\sharp}}=-\frac{1}{R}e^{\ul{4}}, \\
  \omega^{\ul{6}\ul{\sharp}}&=&\frac{1}{R}e^{\ul{7}}, \omega^{\ul{7}\ul{\sharp}}=-\frac{1}{R}e^{\ul{6}},\\
   \omega^{\ul{8}\ul{\sharp}}&=&\frac{1}{R}e^{\ul{9}}, \omega^{\ul{9}\ul{\sharp}}=-\frac{1}{R}e^{\ul{8}}.\eea
And $H_4$ can now be written as
\be H_4=\frac{6}{R} e^{\ul{0}}\w e^{\ul {1}} \w e^{\ul{2}}\w e^{\ul{3}}. \ee

The Killing spinors of $AdS_4\times Q^{1, 1, 1}$ satisfy the following equation
\be \nabla_{\ul{m}}\eta+\frac1{576}(3\Gamma_{{\ul n}{\ul p}{\ul q}{\ul r}}\Gamma_{\ul m}-\Gamma_{\ul m}\Gamma_{{\ul n}{\ul p}{\ul q}{\ul r}})H^{{\ul n}{\ul p}{\ul q}{\ul r}}\eta=0. \ee
Our convention about the product of the eleven $\Gamma$ matrices is \be\Gamma_{\ul{0}\ul{1}\ul{2}\ul{3}\ul{4}\ul{5}\ul{6}\ul{7}\ul{8}\ul{9}\ul{\sharp}}=1.\ee
Using the vielbeins and the spin connections given above, we find that the solution to the above equation is
\be \eta=e^{\frac{u}2\Gamma_{\ul{2}}\hat{\Gamma}}e^{\frac{\rho}2\Gamma_{\ul{1}}\hat{\Gamma}}e^{\frac{t}2\Gamma_{\ul{0}}\hat{\Gamma}}e^{\frac{\phi}2\Gamma_{{\ul{2}}{\ul{3}}}}
\eta_0,\label{spinor} \ee
where $\eta_0$ is independent of all the coordinates 
and satisfies the projection conditions
\be\Gamma^{\ul{4}\ul{5}}\eta_0=\Gamma^{\ul{6}\ul{7}}\eta_0=\Gamma^{\ul{8}\ul{9}}\eta_0,\label{projector1}\ee
and $\hat{\Gamma}$ is defined as
\be \hat{\Gamma}=\Gamma_{\ul{0}\ul{1}\ul{2}\ul{3}}. \ee
The Killing spinors of $Q^{1, 1, 1}$ were also studied in \cite{Hoxha, Donos}. The Killing spinors of $AdS_4$ were given in this coordinate system in \cite{Drukker, drukker4}.

The above projection conditions show that the background on $AdS_4\times Q^{1, 1, 1}$ is $1/4$ BPS, i.e., $8$ supercharges are preserved. These supercharges correspond to $4$ super-Poincare charges and $4$ superconformal charges
in the dual three-dimensional superconformal field theory.

Now we turn to consider the Killing spinors of the orbifolds $AdS_4\times Q^{1, 1, 1}/Z_k$ and
$AdS_4\times Q^{1, 1, 1}/Z_k^\prime$. For this purpose, we compute the Lie-Lorentz derivative of the above Killing spinor $\eta$
with respect to the Killing vector $K_i\equiv\frac{\partial}{\partial\phi_i}$ defined as
\be {\cal L}_{K_i}\eta\equiv(K_i)^{\ul{m}}\nabla_{\ul{m}}\eta+\frac14(\nabla_{\ul{m}}(K_i)_{\ul{n}})\Gamma^{\ul{m}\ul{n}}\eta.\ee
After some calculations, we find \be {\cal L}_{K_i}\eta=0, \ee for each $i$. This result tells us that
 $\eta$ is also the Killing spinor of $AdS_4\times Q^{1, 1, 1}/Z_k$ and
$AdS_4\times Q^{1, 1, 1}/Z_k^\prime$. In other words, the supersymmetries are not broken by these orbifolding.

\section{Probe membrane solutions in $AdS_4\times Q^{1, 1, 1}$}
In this section, we will find two classes of probe $M2$-brane solutions in $AdS_4\times Q^{1, 1, 1}$.
The bosonic part of the $M2$-brane action is:
\be
S_{M2}=T_2\left(\int d^3\xi\sqrt{-\mbox{det}g_{mn}}-\int P[C_3]\right), \ee
where $g_{mn}$ is the induced metric on the membrane, $T_2$ is
the tension of the $M2$-brane: \be T_2={1\over (2\pi)^2l_p^3},\ee and
$P[C_3]$ is the pullback of the bulk 3-form gauge
potential to the worldvolume of the membrane.
The gauge choice for the background 3-form gauge potential  $C_3$ in the case at hand is
\be C_3=\frac{R^3}8(\cosh^3u-1)\cosh\rho dt\w d\rho\w d\phi.
\ee
From the variation of this action, the
membrane equation of motion is \bea
&&\frac{1}{\sqrt{-g}}\p_m\left(\sqrt{-g}g^{mn}\p_nX^{\underline{N}}\right)G_{\underline{MN}}
+g^{mn}\p_{m}X^{\underline{N}}\p_{n}X^{\underline{P}}\G^{\underline{Q}}_{\underline{NP}}G_{\underline{QM}}\nn\\
&=&\frac{1}{3!\sqrt{-g}}\epsilon^{mnp}(P[H_4])_{\underline{M}mnp}.\label{eom}\eea
We always use the indices from the beginning (middle) of the alphabet to refer
to the frame (coordinate) indices, and the underlined indices to
refer to the target space ones. And also notice that $\epsilon^{mnp}$ is a tensor density on the world-volume of the membrane.

We are mainly interested in BPS $M2$-branes. The supersymmetry projector equation reads
\be \Gamma_{M2}\eta=\eta,\ee
with \be\Gamma_{M2}= \frac{1}{\sqrt{-g}}\partial_\tau X^{\mu_1}\partial_\xi X^{\mu_2}\partial_\sigma X^{\mu_3}e^{\ul{m}_1}_{\mu_1}e^{\ul{m}_2}_{\mu_2}e^{\ul{m}_3}_{\mu_3}\Gamma_{\ul{m}_1\ul{m}_2\ul{m}_3}, \ee
where $\tau, \xi, \sigma$ are coordinates on the worldvolume of the $M2$-brane.

\subsection{BPS $M2$-branes dual to Wilson loops revisited}
In this class of solutions, the worldvolume of the $M2$-brane has the topology $AdS_2\times S^1$ with $AdS_2\in AdS_4$
and $S^1\in M_7$. From now on, by $M_7$ we mean either $Q^{1, 1, 1}/Z_k$ or $Q^{1, 1, 1}/Z^\prime_k$. This class
includes $M2$-branes dual to BPS Wilson loops in gauge theories as a special case, and this case was studied in \cite{Sparks}.
In that paper, the authors started with general discussions on BPS Wilson loops in the fundamental representation in ${\cal N}=2$ Chern-Simons-matter theories
and the dual $M2$-brane solutions. They also included $M2$-branes in $AdS_4\times Q^{1, 1, 1}/Z_k$ as one of the explicit examples.
They used a different coordinate system for the $AdS_4$ part and for the $Q^{1, 1, 1}/Z_k$ part they used some results in  differential
geometry which appeared in their general discussions. We will use the explicit results of Killing spinors obtained in the previous section.

The ansatz of these solutions is
\be t=\tau, \rho=\xi, \psi=\psi(\sigma), \phi_i=\phi_i(\sigma), i=1, 2, 3, \ee
with $u, \phi, \theta_i (i=1, 2, 3)$ being constants. Here $\tau, \xi, \sigma$ are three coordinates on the worldvolume of the $M2$-brane.
We consider the case that $\sigma\in [0, 2\pi]$ is a compact direction (i.e. we always identify $\sigma+2\pi$ with $\sigma$).

The periodic conditions for the fields $\psi, \phi_i$ are
\bea \psi(\sigma+2\pi)=\psi(\sigma)+2\pi n_\psi,\\
\phi_1(\sigma+2\pi)=\phi_1(\sigma)+\frac{2\pi n_1}k,\\
\phi_2(\sigma+2\pi)=\phi_2(\sigma)+\frac{2\pi n_1}k+2\pi n_2,\\
\phi_3(\sigma+2\pi)=\phi_3(\sigma)+2\pi n_3,
\eea
with $n_i\in {\bf Z}, i=1, 2, 3$ when $M_7=Q^{1, 1, 1}/Z_k$.

For the case that $M_7=Q^{1, 1, 1}/Z_k^\prime$, the corresponding conditions are
\bea \psi(\sigma+2\pi)=\psi(\sigma)+2\pi n_\psi,\\
\phi_1(\sigma+2\pi)=\phi_1(\sigma)+\frac{2\pi n_1}k,\\
\phi_2(\sigma+2\pi)=\phi_2(\sigma)+2\pi n_2,\\
\phi_3(\sigma+2\pi)=\phi_3(\sigma)+2\pi n_3,
\eea
with $n_i\in {\bf Z}, i=1, 2, 3$.

Now the $M2$-brane action is
\bea S_{M2}&=&\frac{T_{M2}R^3}{4}\int d^3\sigma \cosh^2u\cosh\rho \nn\\
&\times&\left[\frac{1}{8}\sum_{i=1}^3 \sin^2\theta_i \phi_i^{\prime2}+
\frac1{16}(\psi^\prime +\sum_{i=1}^3\cos\theta_i \phi^\prime_i)^2\right]^{1/2},\label{action1}\eea
where $\prime$ means $\partial /\partial\sigma$.
Equation of motion for $u$ gives
\be u=0,\ee
while equation of motion from variation of $\theta_i$ gives
\be \sin\theta_i\phi_i^\prime (\psi^\prime+\sum_{j=1}^3\cos\theta_j\phi_j^\prime-2\cos\theta_i\phi_i^\prime)=0. \label{eom1}\ee
Equations of motion for $\psi, \phi_i$ can be solved by
\be \psi=m_\psi\sigma, \, \phi_i=m_i\sigma.\label{eomsimple}\ee
We also checked that the above three equations are equivalent to the results from the $M2$-brane equations of motion given in
eq.~(\ref{eom}).

To compute the on-shell action of the $M2$-brane whose boundary at infinite is an $S^1$, we switch to the Eclidean $AdS_4$ with the metric:
\be ds^2_4=\frac14(\cosh^2u(d\rho+\sinh^2\rho d\psi^2)+du^2+\sinh^2ud\phi^2).\label{eads}\ee
The on-shell action of the $M2$-brane, eq.~(\ref{action1}), now becomes
\bea S_{M2}&=&\frac{T_{M2}R^3}{4}\int d\Omega_{EAdS_2} d\sigma \nn\\
& &\left[\frac{1}{8}\sum_{i=1}^3 \sin^2\theta_i m_i^2+
\frac1{16}(m_\psi +\sum_{i=1}^3\cos\theta_i m_i)^2\right]^{1/2},\eea
with \be \int d\Omega_{EAdS_2}=\int  d\rho d\psi\sinh \rho.\ee
Using the fact that $\sigma\in[0, 2\pi]$, $T_{M2}=1/(4\pi^2l_p^3)$ and eq.~(\ref{rlp}), we get
\bea S_{M2}&=&\frac{1}{2}\sqrt{\frac{kN}{3}\left(\frac{1}{8}\sum_{i=1}^3 \sin^2\theta_i m_i^2+
\frac1{16}(m_\psi +\sum_{i=1}^3\cos\theta_i m_i)^2\right)}\nn\\
&\times&\int d\Omega_{EAdS_2}.\eea
After adding boundary terms as in \cite{DrukkerGrossOoguri}, we get
\bea S_{M2}=-\pi\sqrt{\frac{kN}{3}\left(\frac{1}{8}\sum_{i=1}^3 \sin^2\theta_i m_i^2+
\frac1{16}(m_\psi +\sum_{i=1}^3\cos\theta_i m_i)^2\right)}.\eea

We now search for BPS $M2$-brane in $AdS_4\times Q^{1, 1, 1}$ among these solutions. $\Gamma_{M2}$ now becomes
\bea \Gamma_{M2}&=&\left(\frac1{16}(\psi^\prime+\sum_{i=1}^3 \cos\theta_i\phi_i^\prime)^2+\frac18 \sum_{i=1}^3\sin^2\theta_i\phi_i^{\prime2}\right)^{-1/2}
\G_{\ul{01}}\nn\\&\times&\left(\frac14 (\psi^\prime+\sum_{i=1}^3\cos\theta_i\phi_i^\prime)\G_{\ul{\sharp}}+
\frac1{\sqrt{2}}\sin\theta_1\phi_1^\prime\G_{\ul{5}}+\frac1{\sqrt{2}}\sin\theta_2\phi_2^\prime\G_{\ul{7}}\right.\nn\\
&-&\left.\frac1{\sqrt{2}}\cos\psi\sin\theta_3\phi_3^\prime\G_{\ul{8}}+\frac1{\sqrt{2}}\sin\psi\sin\theta_3\phi_3^\prime\G_{\ul{9}}\right)\eea
We need that the solutions of $\Gamma_{M2}\eta=\eta$ also satisfy the projection conditions eq.~(\ref{projector1}).
This leads to that for each $i$, \be \sin\theta_i=0, \ee or \be \phi_i^\prime=0. \ee
Now we get
\be\Gamma_{M2}=sign(m_\psi+\sum_{i=1}^3\cos\theta_im_i)\Gamma_{\ul{0}\ul{1}\ul{\sharp}}.\ee
The BPS condition leads to
\be \Gamma_{\ul{0}\ul{1}\ul{\sharp}}\eta=sign(m_\psi+\sum_{i=1}^3\cos\theta_im_i)\eta.\ee
By using the fact that we have $u=0$ on the worldvolume of this $M2$-brane solution, it is not hard to see
that the above condition is equivalent to the condition
\be \Gamma_{\ul{0}\ul{1}\ul{\sharp}}\eta_0=\pm\eta_0,\ee
on the $M2$-brane worldvolume. This condition is compatible with the projection conditions eq.~(\ref{projector1}), and
this BPS $M2$-brane is half-BPS with respect to the background.

The $M2$-brane in $AdS_4\times Q^{1, 1, 1}/Z_k$ dual to half-BPS Wilson loop is a special solution of this class \cite{Sparks}. It is given by
\be m_\psi=0, m_1=m_2=\frac{1}{k}, m_3=0, (\theta_1, \theta_2)=(0, 0), (0, \pi), (\pi, 0), (\pi, \pi).\ee

The result for the on-shell action is
\be S_{M2}=-2\pi\sqrt{\frac{N}{3k}},\ee
when $(\theta_1, \theta_2)=(0, 0), (\pi, \pi),$ while in the case that $(\theta_1, \theta_2)=(0, \pi), (\pi, 0)$
\be S_{M2}=0.\ee

The first two solutions give leading contribution to the vev of Wilson loops, which reads
\be <W>\sim\exp(2\pi\sqrt{\frac{N}{3k}}),\ee
in the leading order of large $N$ expansion. As mentioned in \cite{Sparks}, this is consistent with the result from the matrix model computations in \cite{Nakwoo1}.

Similarly, among the half-BPS $M2$-branes in $AdS_4\times Q^{1, 1, 1}/Z_k^\prime$, the one with
\be m_\psi=0, m_1=\frac1k, m_2=m_3=0, \theta_1=0, \pi\ee
is dual to half-BPS Wilson loops. For the on-shell action
\be S_{M2}=-\pi\sqrt{\frac{N}{3k}},\ee
we get
\be <W>\sim\exp(\pi\sqrt{\frac{N}{3k}}).\ee

\subsection{The second class of solutions}

Now we consider the ansatz
\be t=\tau, \rho=\xi, \phi=\sigma, \ee
\be \psi=\psi(\sigma), \phi_i=\phi_i(\sigma), \ee
with $u, \theta_i$ being constant. We also demand that $u$ is nonzero.
The $M2$-brane action is now
\be S_{M2}=\frac{T_{M2}R^3}{8}\int d^3\sigma \cosh\rho \left[\cosh^2u\sqrt{\sinh^2u+c}-\cosh^3u+1\right],\ee
with  the definition of $c$ \be c\equiv\frac12\sum_{i=1}^3 \sin^2\theta_i^2\phi_i^{\prime2}+\frac14(\psi^\prime+\sum_{i=1}^3\cos\theta_i\phi_i^\prime)^2.\ee
Equation of motion for $u$ gives
\be 2\cosh u\sinh u\sqrt{\sinh^2u+c}+\frac{\cosh^3u \sinh u}{\sqrt{\sinh^2u+c}}-3\sinh u\cosh^2 u=0.\ee

For non-zero $u$, it has two solutions, \be c=1,\ee
and \be c=-\frac34\cosh^2u+1. \ee
From now on we will only consider the first solution which leads to
\be 2\sum_{i=1}^3\sin^2\theta_i\phi_i^{\prime2}+(\psi^\prime+\sum_{i=1}^3\cos\theta_i\phi_i^\prime)^2=4. \ee

Similar to the solutions in the previous subsection,
equation of motion for $\theta_i$ gives
\be \sin\theta_i\phi_i^\prime (\psi^\prime+\sum_{j=1}^3\cos\theta_j\phi_j^\prime-2\cos\theta_i\phi_i^\prime)=0. \ee
And equations of motion for $\psi, \phi_i$ can be solved by
\be \psi=m_\psi\sigma, \, \phi_i=m_i\sigma.\ee
The above  equations are equivalent to the results from the $M2$-brane equations of motion given in
eq.~(\ref{eom}).

Now we turn to discuss the BPS condition for the $M2$-branes in $AdS_4\times Q^{1, 1, 1}$. Now $\Gamma_{M2}$ becomes
\bea \Gamma_{M2}&=&\frac1{\cosh^2u\sqrt{\sinh^2u+c}}\cosh\rho \Gamma_{\ul{0}\ul{1}}(\sinh u\Gamma_{\ul{3}}+\frac12(\psi^\prime+\sum_{i=1}^3\cos\theta_i\phi_i^\prime)\Gamma_{\ul{\sharp}}\nonumber\\
&+&\frac{1}{\sqrt{2}}\sin\theta_1\phi_1^\prime\Gamma_{\ul{5}}+
\frac{1}{\sqrt{2}}\sin\theta_2\phi_2^\prime\Gamma_{\ul{7}}-\frac{1}{\sqrt{2}}\cos\psi\sin\theta_3\phi_3^\prime\Gamma_{\ul{8}}\nonumber\\
&+&
\frac{1}{\sqrt{2}}\sin\psi\sin\theta_3\phi_3^\prime\Gamma_{\ul{9}}).\eea
To have BPS branes, we also need that for each $i$ we have \be  \sin\theta_i=0,\ee
or \be\phi_i^\prime=0.\ee
The fact $c=1$ now leads to
\be \psi^\prime+\sum_{i=1}^3\cos\theta_i\phi_i^\prime=\pm 2. \ee
Using these results, we can
get \be\Gamma_{M2}=\Gamma_{\ul{0}\ul{1}}\left(\frac{\sinh u}{\cosh u}\Gamma_{\ul{3}}\pm\frac1{\cosh u}\Gamma_{\ul{\sharp}}\right).\ee
From eq.~(\ref{spinor}), we can get that \be\Gamma_{M2}\eta=\eta,\ee
is equivalent to \be \Gamma_{\ul{0}\ul{1}\ul{\sharp}}\eta_0=\pm\eta_0.\ee

So when for each $i=1, 2, 3$ we have either $\sin\theta_i=0$ or $\phi_i$ being constant on the worldvolume, the $M2$-branes in this class is half-BPS.
This is similar to the situation in the previous subsection. And after some calculations using the metric in eq.~(\ref{eads}),  we can get that the on-shell action of the $M2$-brane is \be S_{M2}=-2\pi\sqrt{\frac{kN}{3}},\ee
with the boundary term included.

\section{Conclusions and Discussions}
In this paper, we found some BPS $M2$-branes in M-theory on $AdS_4\times Q^{1, 1, 1}$ and its certain orbifolds.
We reproduced the $M2$-branes dual to BPS Wilson loops in the fundamental representation in the field theory side.
We also studied a second class of the BPS $M2$-branes which should include the $M2$-branes dual to vortex loops
in the field theory side. We also find the explicit solution to the Killing spinor equations in this background.

There are several further directions that are interesting for us. For the holographic dual to BPS Wilson loops in the (anti-)fundamental representation,
one should search for suitable $D2$ ($D6$)-brane solutions in the IIA string background obtained from the $S^1$ reduction of  the above M-theory background.
On the other hand, one can try to find suitable $M2$-branes (Kaluza-Klein monopoles) solution in the M-theory background directly.
To correctly identify the dual brane solutions, we also need more precise understanding of the loop operators in the field theory side.
We would also like to try to generalize our studies here to other Sasaki-Einstein $7$-manifolds.
We hope to report our progress in these directions
in the near future.

\section*{Acknowledgments}
The authors are grateful to  Bin Chen, Jarah Evslin, Nakwoo Kim, De-Sheng Li, Wei Li, Feng-Li Lin, Zheng-Wen Liu, Jiang Long, Hong Lu, Jian-Feng Wu,  Gang Yang, Jie Yang and Hossein Yavartanoo
for various helpful discussions. JW would like to thank Lanzhou University for warm hospitality during {\it `Workshop on String/M-theory, Gravity and Topological Field Theories'} and Ehwa Womans University for warm hospitality during {\it `IEU Workshop on Solving AdS/CFT'}.  This work was supported in part by the National Natural
Science Foundation of China under contract No. 11105154 (JW), No. 11222549 (JW), No. 10925522(MZ) and No. 11021092(MZ). JW gratefully acknowledges the support of K.~C.~Wong
Education Foundation and Youth Innovation Promotion Association, CAS
as well.

\end{document}